\documentclass[twocolumn,showpacs,preprintnumbers,amsmath,amssymb,prd]{revtex4}

\newcommand{\gh}{ {\rm gh} \: }
\newcommand{\tr}{ {\rm Tr} \: }
\newcommand{\e}[1]{\epsilon \left( #1 \right) }

\begin{document}
\sloppy
\title{BRST quantization of matrix models with constraints and two-dimensional Yang-Mills theory on the cylinder}
\author{P. V. Buividovich}
\email{buividovich@tut.by}
\affiliation{Belarusian State University, 220080 Belarus, Minsk, Nezalezhnasti av. 4}
\date{January 15, 2007}
\begin{abstract}
BRST quantization of the one-dimensional constrained matrix model which describes two-dimensional Yang-Mills theory on the cylinder is performed. Classical and quantum BRST generators and BRST-invariant hamiltonians are constructed. Evolution operator is expressed in terms of BRST path integral. Advantages of the BRST quantization over the reduced phase space approach leading to the theory of $N$ free fermions are discussed.
\end{abstract}
\maketitle

\section{Introduction}
\label{sec:Introduction}

 Matrix models naturally emerge in string theories, topological field theories and Yang-Mills theories \cite{MakeenkoGaugeMethods, Marino:04:1, Eguchi:82:1}. One of the most popular models which allows one to trace various relations between matrix models, string theories and gauge theories is the two-dimensional pure Yang-Mills theory, which can be interpreted either as a string theory \cite{Gross:93:1, Gross:93:2, Vafa:04} or as a matrix model \cite{Douglas:93:1, Minahan:93:1}.

 Perhaps the most direct way to arrive at the matrix model which describes two-dimensional Yang-Mills theory on the cylinder is to impose the gauge $A_{0} = 0$ and to consider the monodromy of the gauge field around the compactified spatial dimension \cite{Minahan:93:1}:
\begin{eqnarray}
\label{UnitaryMatrixFromYangMills}
W(t) = \mathcal{P} \exp{\left( i g_{YM}^{\:} \int \limits_{0}^{L} dx A_{1}(x,t) \right) }
\end{eqnarray}
where $L$ is the circumference of the compactified spatial dimension and $g_{YM}^{\:}$ is the coupling constant of the Yang-Mills theory. The Yang-Mills lagrangian can be then expressed in terms of $W(t)$ as \cite{Minahan:93:1}:
\begin{eqnarray}
\label{2DYMSphereMMLagrangian}
\mathcal{L} = \frac{1}{2} \int \limits_{0}^{L} dx \tr F_{01}^{2} = - \frac{1}{2 g^{2}_{YM} L} \: \tr \left( W^{-1} \dot{W} \right)^{2}
\end{eqnarray}
which is the lagrangian for the one-dimensional $c=1$ matrix model \cite{Douglas:93:1, Minahan:93:1}. In addition, (\ref{UnitaryMatrixFromYangMills}) implies the following constraints \cite{Minahan:93:1}:
\begin{eqnarray}
\label{2DYMTorusMMConstraint}
\left[W, \dot{W} \right] = 0
\end{eqnarray}
If the boundaries of the cylinder are situated at $t = t_{1}$ and $t = t_{2}$, the definition (\ref{UnitaryMatrixFromYangMills}) implies that $W(t_{1})$ and $W(t_{2})$ are the holonomies of the gauge field around these boundaries.

The constraints (\ref{2DYMTorusMMConstraint}) are first-class constraints for the lagrangian (\ref{2DYMSphereMMLagrangian}), therefore one can apply the usual quantization methods developed for systems with first-class constraints \cite{HenneauxQuantGauge}, such as implementing reduced phase space variables, introducing new second-class constraints or imposing BRST symmetry on the extended phase space complemented with ghost variables \cite{HenneauxQuantGauge}. The approach of  \cite{Douglas:93:1, Minahan:93:1} is exactly the reduced phase space approach. Fortunately, for the constraints (\ref{2DYMTorusMMConstraint}) the reduced phase space variables are simply the eigenvalues of the matrix $W$ and the reduced hamiltonian is explicitly known - the model is equivalent to $N$ free nonrelativistic fermions \cite{Douglas:93:1, Minahan:93:1} and can be investigated using the methods of conformal field theory.

 However, very often it is quite difficult to find the full set of reduced phase space variables. For instance, in the case of models with several matrix-valued variables it is impossible to transform all matrices to diagonal form simultaneously, and the methods developed in \cite{Douglas:93:1, Minahan:93:1} are not applicable. Furthermore, it is often desirable to preserve the covariance of the theory in terms of the original variables. For example, in the context of correspondence between string theories and matrix models, or for two-dimensional Yang-Mills theory, it could make the identification of stringy degrees of freedom easier \cite{Marino:04:1, MakeenkoGaugeMethods}. Preserving the group structure can be also advantageous if one wants to use finite-dimensional Lie groups as regularizations of infinite-dimensional groups such as the group of area-preserving diffeomorphisms $SDiff(\mathcal{M})$ arising in the description of relativistic membranes in the light-cone gauge \cite{Pope:90:1, Hoppe:88}. In these cases BRST quantization is more suitable.

The aim of this paper is to perform BRST quantization of the matrix model (\ref{2DYMSphereMMLagrangian}) with the constraints (\ref{2DYMTorusMMConstraint}). In order to make the analysis as general as possible the gauge group will not be specified and the coordinates on the group manifold will not be fixed. Only local features of the group geometry will be considered. Hopefully such general analysis can be extended to the models with several matrix-valued variables and to infinite-dimensional groups relevant in string theories.

The structure of this paper is the following: in the section \ref{sec:Geometry} the basic concepts of classical and quantum mechanics on the group manifold are reviewed and the constraints (\ref{2DYMTorusMMConstraint}) are shown to be first-class and reducible. Hamiltonian formalism is used, as the BRST construction is inherent to canonical quantization \cite{HenneauxQuantGauge}. Classical and quantum BRST generators for the constraints (\ref{2DYMTorusMMConstraint}) and BRST-invariant hamiltonians are constructed in the sections \ref{sec:ClassicalBRST} and \ref{sec:QuantumBRST} respectively. BRST path integral for the lagrangian (\ref{2DYMSphereMMLagrangian}) is considered in the section \ref{sec:PathInt}. Some technical details such as explicit expressions for the geometric constructions used throughout the paper are relegated to the appendices.

\section{Classical and quantum mechanics on the group manifold}
\label{sec:Geometry}

 In this section classical and quantum mechanics on the group manifold are briefly reviewed. Group generators in the fundamental representation are denoted as $T_{a}$. The structure constants $C^{c}_{ab}$ and the metrics $c_{a b}$ of the Lie algebra are fixed by the relations:
\begin{eqnarray}
\label{LieAlgebraProperties}
\tr \left( T_{a} T_{b} \right) = c_{a b}, \quad
[T_{a}, T_{b}] = i C^{c}_{ab} T_{c}
\end{eqnarray}

  Suppose that group elements are parametrized by some coordinates $x^{\alpha}$. From now on latin indices $a, b, c, \ldots$ will be used to denote the elements of the Lie algebra and the first greek indices $\alpha, \beta, \ldots$ for tensors on the group manifold. Left and right actions of the group generators $T_{a}$ define the left- and right-invariant vector fields $L^{\alpha}_{a}(x)$ and $R^{\alpha}_{a}(x)$ \cite{DeWittGroupsAndFields}:
\begin{eqnarray}
\label{LeftRightVectorFields}
L^{\alpha}_{a} \partial_{\alpha} W(x) = i T_{a} W(x), \quad  R^{\alpha}_{a} \partial_{\alpha} W(x) = i W(x) T_{a}
\end{eqnarray}
where $W(x)$ are the matrices of the fundamental representation of the group. The difference of the vector fields $L^{\alpha}_{a}$ and $R^{\alpha}_{a}$, which will be denoted as $Q^{\alpha}_{a} = L^{\alpha}_{a} - R^{\alpha}_{a}$, generates shifts within group classes:
\begin{eqnarray}
\label{ClassVectorFields}
Q^{\alpha}_{a} \partial_{\alpha} W(x) = i \left[ T_{a}, W(x) \right]
\end{eqnarray}
It is important to note that while the vector fields $L_{a}^{\alpha}$ and $R_{a}^{\alpha}$ build complete bases in each point, the vector fields $Q^{\alpha}_{a}$ are not all independent locally, but build an overfull basis in the tangent space to the group classes, which means that there exist such functions $u^{a}_{k}$ that $u^{a}_{k} Q_{a}^{\alpha} = 0$. One can show that this condition is equivalent to the orthogonality of the vector fields $u_{k}^{\alpha} = u^{a}_{k} L_{a}^{\alpha}$ and $Q^{\alpha}_{a}$: $g_{\alpha \beta} Q_{a}^{\alpha} u^{\beta}_{k} = 0$. It will be assumed that $u_{k}^{a}$ are linearly independent reducibility conditions, therefore vector fields $u_{k}^{\alpha}$ and $Q_{a}^{\alpha}$ together build an overfull basis in the tangent space at each point of the group manifold.

It is also convenient to introduce the left- and right- invariant 1-forms $L_{\alpha}^{a}$ and $R_{\alpha}^{a}$, defined as inverse to the vector fields $L^{\alpha}_{a}$ and $R^{\alpha}_{a}$ respectively:
\begin{eqnarray}
\label{LRFormsDef}
L_{\alpha}^{a} L^{\beta}_{a} = \delta_{\alpha}^{\beta}, \quad R_{\alpha}^{a} R^{\beta}_{a} = \delta_{\alpha}^{\beta}
\end{eqnarray}
These 1-forms take especially simple form in matrix notation:
\begin{eqnarray}
\label{LRFormsMatrix}
L_{\alpha} = L_{\alpha}^{a} T_{a} =  - i \partial_{\alpha} W W^{-1}, \nonumber \\
R_{\alpha} = R_{\alpha}^{a} T_{a} = - i W^{-1} \partial_{\alpha} W
\end{eqnarray}

The lagrangian (\ref{2DYMSphereMMLagrangian}) can now be rewritten in terms of the group metric (Killing form) $g_{\alpha \beta} = c_{ab} L^{a}_{\alpha} L^{b}_{\beta} = c_{ab} R^{a}_{\alpha} R^{b}_{\beta} = \tr \left( L_{\alpha} L_{\beta} \right) = \tr \left( R_{\alpha} R_{\beta} \right)$ as:
\begin{eqnarray}
\label{2DYMSphereMMLagrangianGeo}
\mathcal{L} = \frac{1}{2 g^{2}_{YM} L} \: g_{\alpha \beta} \dot{x}^{\alpha} \dot{x}^{\beta}
\end{eqnarray}
while the constraints (\ref{2DYMTorusMMConstraint}) can be written as $Q^{a}_{\alpha} \dot{x}^{\alpha} = 0$.

In hamiltonian formalism the transformations (\ref{LeftRightVectorFields}) and (\ref{ClassVectorFields}) are generated by the functions $L_{a} = L_{a}^{\alpha} p_{\alpha}$, $R_{a} = R_{a}^{\alpha} p_{\alpha}$ and $Q_{a} = L_{a} - R_{a}$, where $p_{\alpha}$ are the momenta canonically conjugate to $x^{\alpha}$: $\{ x^{\alpha}, p_{\beta}  \} = \delta^{\alpha}_{\beta}$. The algebra of these functions is the Lie algebra of the group:
\begin{eqnarray}
\label{CommutationRelationsClassic}
\{ L_{a}, L_{b} \} = - C^{c}_{a b} L_{c}, \quad \{ R_{a}, R_{b} \} = C^{c}_{a b} R_{c}  \nonumber \\
\{ L_{a}, R_{b} \} = 0, \quad \{ Q_{a}, Q_{b} \} = - C^{c}_{a b} Q_{c}
\end{eqnarray}

The hamiltonian corresponding to (\ref{2DYMSphereMMLagrangian}) and (\ref{2DYMSphereMMLagrangianGeo}) is:
\begin{eqnarray}
\label{InvariantHamiltonianClassic}
H_{0} = \frac{g^{2}_{YM} L}{2} \: c^{ab} L_{a} L_{b} = \nonumber \\ =
 \frac{g^{2}_{YM} L}{2} \: c^{ab} R_{a} R_{b} = \frac{g^{2}_{YM} L}{2} \: g^{\alpha \beta} p_{\alpha} p_{\beta}
\end{eqnarray}
while the constraints (\ref{2DYMTorusMMConstraint}) are simply $Q_{a} = 0$. As $Q_{a}$ are the integrals of motion for the hamiltonian (\ref{InvariantHamiltonianClassic}) and build the closed algebra (\ref{CommutationRelationsClassic}), the constraints (\ref{2DYMTorusMMConstraint}) are indeed first-class.

In quantum mechanics the counterparts of the classical functions $L_{a}$, $R_{a}$ and $Q_{a}$ are the differential operators $\hat{L}_{a} = - i L_{a}^{\alpha} \partial_{\alpha}$, $\hat{R}_{a} =  - i R_{a}^{\alpha} \partial_{\alpha}$ and $\hat{Q}_{a} = \hat{L}_{a} - \hat{R}_{a}$, which also build a representation of the Lie algebra (\ref{LieAlgebraProperties}):
\begin{eqnarray}
\label{CommutationRelationsQuantum}
\left[ \hat{L}_{a}, \hat{L}_{b} \right] = - i C^{c}_{a b} \hat{L}_{c}, \quad \left[ \hat{R}_{a}, \hat{R}_{b} \right] = i C^{c}_{a b} \hat{R}_{c} \nonumber \\
\left[ \hat{L}_{a}, \hat{R}_{b} \right] = 0, \quad \left[ \hat{Q}_{a}, \hat{Q}_{b} \right] = - i C^{c}_{a b} \hat{Q}_{c}
\end{eqnarray}

Quantum counterpart of $H_{0}$ is proportional to the Laplace operator on the group, or the second-order Casimir operator on the space of differentiable functions on the group manifold:
\begin{eqnarray}
\label{InvariantHamiltonianQuantum}
\hat{H}_{0} = \frac{g^{2}_{YM} L}{2}  \hat{L}_{a} \hat{L}_{a} = \nonumber \\ =
- \frac{g^{2}_{YM} L}{2} \:  \sqrt{g^{-1}} \partial_{\alpha} \left( \sqrt{g} g^{\alpha \beta} \partial_{\beta} \right) \quad
\end{eqnarray}
where $g = {\rm det} \: (g_{\alpha \beta})$. In the case of compact groups eigenvalues of $\hat{H}_{0}$ are proportional to the eigenvalues of quadratic Casimir operators of irreducible unitary representations of the group and the corresponding eigenfunctions are the matrices of these representations \cite{HoweTanNonAbelianHarmonicAnalysis}. For the sake of brevity from now the factor $\frac{g^{2}_{YM} L}{2}$ in front of the hamiltonian will be omitted.

\section{Classical BRST generator and BRST-invariant hamiltonian}
\label{sec:ClassicalBRST}

The constraints $Q_{a} = 0$ on the original phase space can be replaced by the requirement of invariance under BRST transformation generated by the BRST generator $\Omega$ on the extended phase space complemented with ghost variables \cite{HenneauxQuantGauge}. In order to do this, two generations of ghosts should be introduced: fermionic ghosts $X^{a}$ and the conjugate ghost momenta $\Pi_{a}$ which correspond to the constraints $Q_{a} = 0$, and bosonic "ghosts of ghosts" $\chi^{k}$ and the conjugate momenta $\pi_{k}$ which are associated with the  reducibility conditions $u^{a}_{k} Q_{a} = 0$ \cite{Henneaux:93, HenneauxQuantGauge}. Ghost numbers of these variables are \cite{HenneauxQuantGauge}:
\begin{eqnarray}
\label{GhostNumbers}
\gh X^{a} = 1, \: \: \gh \Pi_{a} = -1, \: \:  \gh \chi^{k} = 2, \: \: \gh \pi_{k} = - 2
\end{eqnarray}

Any BRST generator should be nilpotent, i.e. $\{ \Omega, \Omega \} = 0$, and be of ghost number one. In order to define a proper BRST cohomology, BRST generator should have the form $\Omega = X^{a} Q_{a} + \chi^{k} u_{k}^{a} \Pi_{a} + \: (more)$, where $(more)$ contains higher-order ghost terms \cite{HenneauxQuantGauge}. The first guess is to combine this expression with the simplest possible expression for reducible first-class constraints which build a closed algebra \cite{HenneauxQuantGauge, Henneaux:93}:
\begin{eqnarray}
\label{BRSTChargeClassic}
\Omega = X^{a} Q_{a} + \chi^{k} u_{k}^{a} \Pi_{a} - 1/2 \: C^{c}_{a b} X^{a} X^{b} \Pi_{c}
\end{eqnarray}
The nilpotency condition reads explicitly:
\begin{eqnarray}
\label{Nilpotency}
1/2 \{ \Omega, \Omega \} = \frac{\partial \Omega}{\partial x^{\alpha}} \: \frac{\partial \Omega}{\partial p_{\alpha}} +
\frac{\partial \Omega}{\partial \chi^{k}} \: \frac{\partial \Omega}{\partial \pi_{k}} -
\frac{\partial^{L} \Omega}{\partial X^{a}} \: \frac{\partial^{L} \Omega}{\partial \Pi_{a}} = \nonumber \\ =
\chi^{k} \Pi_{a} X^{b} \left( \{u^{a}_{k}, Q_{b} \} - C^{a}_{f b} u^{f}_{k} \right) = 0 \quad
\end{eqnarray}
Thus the BRST generator of the form (\ref{BRSTChargeClassic}) is nilpotent iff $\{u^{a}_{k}, Q_{b} \} - C^{a}_{f b} u^{f}_{k} = 0$. This equation can be formulated as the commutativity of the vector fields $u_{k} = u^{a}_{k} L_{a}$ and $Q_{a}$:
\begin{eqnarray}
\label{NilpotencyCondition}
\{Q_{a}, u_{k} \} = 0
\end{eqnarray}

It follows from (\ref{Nilpotency}) that $u^{a}_{k}$ span the Cartan subspace of the Lie algebra (however, different Cartan subspace in each point), since $u^{a}_{k} u^{b}_{l} C_{a b}^{c} = 0$. The vector fields $u^{\alpha}_{k}$ which satisfy the commutation relations (\ref{NilpotencyCondition}) are explicitly constructed for $SU(N)$ group in Appendix \ref{sec:SUNPolar}. This construction can be easily generalized for an arbitrary Lie group.

A function $F$ on the extended phase space is called BRST-closed if $\{ \Omega, F \} = 0$ and BRST-exact if it is equal to the Poisson brackets of some other function with $\Omega$: $F = \{ \Omega, G \}$. Nilpotent BRST generator defines the BRST cohomology on the space of functions on the extended phase space, with cohomological classes consisting of BRST-closed functions modulo BRST-exact functions \cite{HenneauxQuantGauge}. A proper BRST-generator is defined in such a way that the only nontrivial cohomological class contains gauge-invariant functions. The hamiltonian $H_{0}$ is a gauge-invariant function on the original phase space, but not necessarily on the extended phase space, as in general $\{\Omega, H_{0} \} \neq 0$. In order to construct an invariant hamiltonian on the extended phase space, one should add to $H_{0}$ some terms which contain ghost variables. A proper BRST-invariant extension $H$ of the hamiltonian $H_{0}$ should have ghost number zero and even grassman parity, be BRST-closed and be equal to $H_{0}$ if all ghost variables are set to zero. The ghost number zero  and grassman parity conditions are restrictive enough to eliminate many possible ghost terms. The first nontrivial extension of $H_{0}$ with the maximal antighost number $2$ is:
\begin{eqnarray}
\label{BRSTHamiltonianClassic}
H = H_{0} + \chi^{k} h^{a b}_{k} \Pi_{a} \Pi_{b}
\end{eqnarray}
where $h^{a b}_{k}$ is antisymmetric in $a$ and $b$ and depends only on $x^{\alpha}$. Poisson brackets of the hamiltonian $H$ and the BRST generator (\ref{BRSTChargeClassic}) are:
\begin{eqnarray}
\label{OmegaHPoissonBrackets}
\{ \Omega, H \} =
\{\Omega, H  \}_{pq} + \{ \Omega, H \}_{\pi \chi} - \nonumber \\
 - \frac{\partial^{L} H}{\partial \Pi_{a}} \: \frac{\partial^{L} \Omega}{\partial X^{a}}
 - \frac{\partial^{L} H}{\partial X^{a}} \: \frac{\partial^{L} \Omega}{\partial \Pi_{a}} = \nonumber \\ =
\chi^{k} \Pi_{a} \left( \{ u^{a}_{k},H_{0} \} - 2 h^{b a}_{k} Q_{b} \right) + \nonumber \\ +
\chi^{k} \Pi_{a} \Pi_{b} X^{c} \left( \{Q_{c}, h^{a b}_{k}\} - 2 h^{d a}_{k} C^{b}_{d c}  \right)
\end{eqnarray}
where $\{ \Omega , H \}_{pq}$ and $\{ \Omega , H \}_{\pi \chi}$ are Poisson brackets w.r.t. the variables $\left( x^{\alpha}, \: p_{\alpha} \right)$ and $\left( \chi^{k}, \: \pi_{k} \right)$. Thus the hamiltonian $H$ is BRST closed iff the following equations hold:
\begin{eqnarray}
\label{ClosednessCondition}
\{u^{a}_{k}, H_{0} \} + 2 h^{a b}_{k} Q_{b} = 0 \nonumber \\
\{Q_{c}, h^{a b}_{k} \} - h^{d [ a}_{k} C^{b ]}_{d c} = 0
\end{eqnarray}

These equations are solved in Appendix \ref{sec:ClosednessSolution}. It turns out that the solution is most conveniently represented in terms of the 2-form $h_{k \: \alpha \beta} = h^{a b}_{k} Q_{a \: \alpha} Q_{b \: \beta}$:
\begin{equation}
\label{ClosednessSolution}
h_{k} = -1/2 \: u^{a}_{k} \: d Q_{a}, \: \:
h_{k \: \alpha \beta} = - i \tr \left( u_{k} \left[ L_{\alpha}, L_{\beta} \right] \right)
\end{equation}
where $d Q_{a}$ is the external derivative of the 1-form $Q_{a \: \alpha} = g_{\alpha \beta} Q^{\beta}_{a}$ and $u_{k} = u^{a}_{k} T_{a}$. If $S^{b \: \beta}$ are such vector fields that $Q_{a \: \alpha} S^{b \: \alpha} = P^{a}_{b}$, where $P^{a}_{b}$ is the projective operator which projects on the subspace in Lie algebra spanned on $Q_{a}^{\alpha}$, than $h_{k}^{a b}$ can be obtained as:
\begin{eqnarray}
\label{ClosednessSolution1}
h_{k}^{a b} = h_{k \: \alpha \beta} S^{a \: \alpha} S^{b \: \beta}
\end{eqnarray}
The vector fields $S^{a \: \alpha}$ are explicitly constructed for $SU(N)$ group in the Appendix \ref{sec:SUNPolar}. Thus the BRST-invariant extension of the hamiltonian $H_{0}$ has been constructed.

It is also interesting to note that the first equation in (\ref{ClosednessCondition}) implies the commutativity of the vector fields $u^{\alpha}_{k}$:
\begin{equation}
\label{uCommutation}
\{u_{k}, u_{l} \} = 0
\end{equation}

The BRST-invariant hamiltonian (\ref{BRSTHamiltonianClassic}) is still quadratic in the momenta and thus can be thought of as a free hamiltonian on a supermanifold with coordinates $x^{\alpha}, X^{a}, \chi^{k}$.

\section{Quantum BRST generator and BRST-invariant hamiltonian}
\label{sec:QuantumBRST}

There are several methods to quantize theories with first-class constraints: quantization on the reduced phase space, Dirac quantization and BRST quantization. Reduced phase space quantization amounts to eliminating all nondynamical coordinates and dealing only with dynamical ones. For the matrix model (\ref{2DYMSphereMMLagrangian}) with the constraints (\ref{2DYMTorusMMConstraint}) the reduced Hilbert space is equivalent to the Hilbert space of $N$ nonrelativistic free fermions \cite{Minahan:93:1, Douglas:93:1}. In the Dirac quantization method the constraints are imposed on the states of the Hilbert space of the unconstrained system:
\begin{eqnarray}
\label{ConstraintQuantum}
\hat{Q}_{a} | \phi \rangle = 0
\end{eqnarray}
The constraints (\ref{ConstraintQuantum}) are also first-class quantum-mechanically, as $[\hat{Q}_{a}, \hat{H}_{0}] = 0$ and $\hat{Q}_{a}$ build the algebra (\ref{CommutationRelationsQuantum}).

The Hilbert space of the unconstrained system (\ref{2DYMSphereMMLagrangian}) is the space of functions on the group manifold. In this paper wave functions in the Schr\"{o}dinger representation are assumed to be scalar functions (not 1/2-densities, as was suggested in the work \cite{Henneaux:93} and also implicitly assumed in the works \cite{Minahan:93:1, Douglas:93:1}). When the group is simple and the equation (\ref{NilpotencyCondition}) holds for the constraints and reducibility conditions, scalar wave functions differ from those of \cite{Henneaux:93} just by the factor $\sqrt{g}$. In the case of compact groups the matrices of irreducible unitary representations build an orthonormal basis in the space of functions on the group manifold \cite{HoweTanNonAbelianHarmonicAnalysis}. The constraints $\hat{Q}_{a} | \phi \rangle = 0$ select the subspace of functions on the group classes. Group characters build an orthonormal basis in this subspace \cite{HoweTanNonAbelianHarmonicAnalysis}.

In the BRST quantization method the constraints $\hat{Q}_{a} | \phi \rangle = 0$ are replaced by a single constraint $\hat{\Omega} | \Phi \rangle = 0$ on the extended Hilbert space complemented with ghost states, where $\hat{\Omega}$ is the quantum counterpart of the classical BRST generator \cite{HenneauxQuantGauge, Henneaux:93, Henneaux:94}. In this paper the Schr\"{o}dinger representation will be used, where wave functions depend on the original variables $x^{\alpha}$ as well as on the ghost variables $\chi^{k}$ and $X^{a}$. The ghost momenta operators are $\hat{\Pi}_{a} = -i \frac{\partial^{L}}{\partial X^{a}}$ and $\hat{\pi}_{k} = - i \frac{\partial}{\partial \chi^{k}}$. Any quantum BRST generator $\hat{\Omega}$ should be nilpotent, self-conjugate and of ghost number one. As $\hat{\Omega}$ is nilpotent, one can define the quantum BRST cohomology as the quotient space of BRST-closed states modulo BRST-exact states. For a properly constructed BRST generator the only nontrivial cohomological class contains physical states. The results of \cite{HenneauxQuantGauge, Henneaux:93, Henneaux:94} imply that for the hamiltonian (\ref{InvariantHamiltonianQuantum}) the proper quantum BRST generator for the constraints (\ref{ConstraintQuantum}) and the corresponding BRST-invariant hamiltonian can be directly obtained from the classical expressions (\ref{BRSTChargeClassic}) and (\ref{BRSTHamiltonianClassic}):
\begin{eqnarray}
\label{BRSTChargeQuantum}
\hat{\Omega} = \hat{X}^{a} \hat{Q}_{a} + \hat{\chi}^{k} u_{k}^{a} \hat{\Pi}_{a} - 1/2 \: C^{c}_{a b} \hat{X}^{a} \hat{X}^{b} \hat{\Pi}_{c}
\end{eqnarray}
\begin{eqnarray}
\label{BRSTHamiltonianQuantum}
\hat{H} = \hat{H}_{0} + \hat{\chi}^{k} h^{a b}_{k} \hat{\Pi}_{a} \hat{\Pi}_{b}
\end{eqnarray}
No operator-ordering ambiguities arise for such BRST generator and BRST-invariant hamiltonian in the case of simple groups.

\section{BRST path integral}
\label{sec:PathInt}

In this section the BRST path integral representation for the evolution operator of the model (\ref{2DYMSphereMMLagrangian}) with the constraint (\ref{2DYMTorusMMConstraint}) will be constructed. In terms of two-dimensional Yang-Mills theory on the cylinder this evolution operator is the Wick-rotated partition function.

In the Dirac quantization method the constraints (\ref{ConstraintQuantum}) select the subspace of singlet states. Therefore the kernel of the evolution operator of the constrained matrix model (\ref{2DYMSphereMMLagrangian}) can be written as (the factor $\left( 2 g^{2}_{YM} L \right)$ in front of the hamiltonian is again omitted for the sake of brevity) \cite{Minahan:93:1, Douglas:93:1}:
\begin{eqnarray}
\label{EvolutionOperator}
U \left( W(t_{1}), W(t_{2}) \right) = \nonumber \\ =
\sum \limits_{R} \bar{\chi}_{R} \left(W(t_{2}) \right) \chi_{R} \left(W(t_{1}) \right)
\exp{\left( - i (t_{2} - t_{1}) C_{2 \: R} \right)} = \nonumber \\ =
\langle W(t_{2}) | \hat{P}_{s} \exp{\left( - i \hat{H}_{0} (t_{2} - t_{1}) \right)} \hat{P}_{s} | W(t_{1})  \rangle \quad
\end{eqnarray}
where $\chi_{R}$ and $C_{2 \: R}$ are the group characters and the second-order Casimir in the representation $R$ and the operator $\hat{P}_{s}$ projects on the subspace of functions on the group classes. The same kernel can be written as the so-called projected kernel in the Dirac quantization procedure \cite{HenneauxQuantGauge}, which is obtained by inserting the operator version of the gauge-fixing conditions between bras and kets of the initial and final physical states \cite{HenneauxQuantGauge}. In the Schr\"{o}dinger representation this amounts to eliminating the volume of the gauge orbits from the integrals over configuration space of the system. However, in order to recover the partition function (\ref{EvolutionOperator}) from the projected kernel one should use the scalar product in the Hilbert space of the unconstrained system, i.e. $\langle \phi_{1} | \phi_{2} \rangle = \int d^{n} x \sqrt{g} \: \bar{\phi}_{1} \phi_{2}$, where $d^{n} x \sqrt{g}$  is the invariant measure on the group manifold and $n$ is the dimensionality of the group. Group characters are orthonormal w.r.t. this scalar product, and therefore all the states will be counted with the same weight and the expression (\ref{EvolutionOperator}) will be reproduced.

In order to establish a mathematically precise connection between the projected kernel and the kernel of the BRST-invariant extension of the operator, the phase space should be further enlarged by the variables of the so-called nonminimal sector \cite{HenneauxQuantGauge, Henneaux:93, Henneaux:94}. These variables are the lagrange multipliers $\lambda^{s \: a_{s}}_{k}$ and the conjugate momenta $b^{k}_{s \: a_{s}}$ as well as the ghosts $C^{s}_{k \: a _{s}}$ and ghost momenta $\rho^{k \: a_{s}}_{s}$ which correspond to the first-class constraints $b^{k}_{s \: a_{s}} = 0$, where $0 \le k \le 1$, $k \le s \le 1$ \cite{Henneaux:94}. The indices $a_{0}, \: b_{0}, \: c_{0}, \: \ldots$ with subscript zero are equivalent to the latin indices $a,\: b,\: c$ and the indices $a_{1}, \: b_{1}, \: c_{1}$ are equivalent to the indices $k, \: l, \: m, \: \ldots$ labelling reducibility conditions. Poisson brackets of the variables of the nonminimal sector are \cite{Henneaux:94}:
\begin{eqnarray}
\label{NonminimalCommutators}
\{ b^{k'}_{s' \: b_{s'}}, \lambda^{s \: a_{s}}_{k}  \} = \{ \rho^{k' \: a_{s}}_{s}, C^{s'}_{k \: b_{s'}} \} = - \delta^{k'}_{k} \delta^{s}_{s'} \delta^{a_{s}}_{b_{s'}}
\end{eqnarray}
Grassman parities and ghost numbers of the variables of the nonminimal sector are \cite{Henneaux:94}:
\begin{eqnarray}
\label{NonminimalGhostNumbers}
\e{b^{k}_{s \: a_{s}}} = \e{\lambda^{s \: a_{s}}_{k}} = s - k \mod 2 \nonumber \\
\e{\rho^{k \: a_{s}}_{s}} = \e{C^{s}_{k \: a_{s}}} = s - k + 1 \mod 2 \nonumber \\
\gh b^{k}_{s \: a_{s}} = - \gh \lambda^{s \: a_{s}}_{k} = k - s \nonumber \\
\gh \rho^{k \: a_{s}}_{s} = - \gh C^{s}_{k \: a_{s}} = s - k + 1
\end{eqnarray}
The variables of the nonminimal sector should be also included in the quantum and classical BRST generators (\ref{BRSTChargeQuantum}) and (\ref{BRSTChargeClassic}). Additional terms which should be added to (\ref{BRSTChargeQuantum}) and (\ref{BRSTChargeClassic}) have the following form \cite{Henneaux:94}:
\begin{eqnarray}
\label{BRSTChargeNonminimal}
\hat{\Omega}_{nonmin} = \sum \limits_{k = 0}^{1} \sum \limits_{s = k}^{1} \hat{b}^{k}_{s \: a_{s}} \hat{\rho}^{k \: a_{s}}_{s}
\end{eqnarray}

It was proven in \cite{Henneaux:94, HenneauxQuantGauge} that the projected kernel of a gauge-invariant operator on the Hilbert space of the unconstrained system is equal to the kernel of the BRST-invariant extension of this operator with suitable ghost states and a suitable gauge-fixing fermion $\hat{K}$. As the evolution operator is a gauge-invariant operator, the projected kernel $U \left( W(t_{1}), W(t_{2}) \right)$ can be expressed in terms of the BRST-invariant extension (\ref{BRSTHamiltonianQuantum}) of the hamiltonian $\hat{H}_{0}$:
\begin{widetext}
\begin{eqnarray}
\label{ProjectedKernelAndBRST}
U \left( W(t_{1}), W(t_{2}) \right) =
 \langle W(t_{2}) | \hat{P}_{s}
 \exp{\left( - i \hat{H}_{0} (t_{2} - t_{1}) \right)}
 \hat{P}_{s} | W(t_{1}) \rangle
 = \nonumber \\ =
\langle W(t_{2}) | \langle ghosts | \hat{P}_{s}
 \exp{\left( - i \left( \hat{H} - [\hat{K}, \hat{\Omega}] \right) (t_{2} - t_{1}) \right)}
 \hat{P}_{s} | W(t_{1}) \rangle | ghosts \rangle
\end{eqnarray}
\end{widetext}
The states $| W \rangle$ are the eigenstates of the coordinate operator and are normalized to the delta-function on the group manifold. In order to get rid of the projection operators $\hat{P}_{s}$ one should consider the states $| \tilde{W} \rangle = \int dV \: | V W V^{-1} \rangle $ smeared over the gauge orbits, i.e. over the group classes. Ghost states, denoted as $|ghosts \rangle$ in (\ref{ProjectedKernelAndBRST}), are the eigenstates of ghost variables with zero eigenvalues \cite{HenneauxQuantGauge, Henneaux:94}:
\begin{eqnarray}
\label{GhostBoundaryConditions}
\hat{X}^{a} | ghosts \rangle = 0, \quad \hat{\chi}^{k} | ghosts \rangle = 0 \nonumber \\
\hat{b}^{k}_{s \: a_{s}} | ghosts \rangle = 0, \quad \hat{C}^{s}_{k \: a_{s}} | ghosts \rangle = 0, \: k \: even \nonumber \\
\hat{\lambda}^{s \: a_{s}}_{k} | ghosts \rangle = 0, \quad \hat{\rho}^{k \: a_{s}}_{s} | ghosts \rangle = 0, \: k \: odd
\end{eqnarray}

The gauge-fixing  fermion $\hat{K}$ in (\ref{ProjectedKernelAndBRST}) regularizes the products of physical states and mixes the variables of the minimal and nonminimal sectors, the former property being important for the existence of BRST cohomology at ghost number zero. A careful examination of the proof of the theorem 14.9 in \cite{HenneauxQuantGauge} leads to the conclusion that the additional terms to be included in BRST extensions of the kernels of gauge-invariant operators are:
\begin{eqnarray}
\label{GFFermion}
\exp{\left( i\left[ \hat{K}, \hat{\Omega} \right] \right)} = \exp{ \left( i \hat{\lambda}^{0 a}_{0} \hat{Q}_{a} + i \hat{\Pi}_{a} \hat{\rho}^{0 a}_{0}  - i \hat{\pi}_{k} \hat{\rho}^{0 k}_{1} \right) }
\end{eqnarray}

Following the proof of the theorem 3 in \cite{Henneaux:94} and using the standard time slicing procedure one can express the projected kernel of the evolution operator (\ref{EvolutionOperator}) in terms of the path integral:
\begin{widetext}
\begin{eqnarray}
\label{EvolutionOperatorPathIntBRST}
U( W(t_{1}), W(t_{2}) ) = \int \limits_{W(t_{1})}^{W(t_{2})} \mathcal{D} x^{A} \mathcal{D} p_{A}
\: \exp{ \left( i \int \limits_{t_{1}}^{t_{2}} d t \left( p_{A} \dot{x}^{A}  - H  +  \lambda^{0 a}_{0} Q_{a} + \Pi_{a} \rho^{0 a}_{0}  - \pi_{k} \rho^{0 k}_{1} \right)
\right)}
\end{eqnarray}
\end{widetext}
where $x^{A}$ denotes all coordinates, including ghosts and the nonminimal sector, and $p_{A}$ all conjugate momenta. Boundary conditions for the ghost variables and for the variables of the nonminmal sector are given by (\ref{GhostBoundaryConditions}).

The momenta variables can be integrated out in order to see how the lagrangian (\ref{2DYMSphereMMLagrangian}) is modified, however, after that the explicit BRST symmetry is lost. It is also more convenient to integrate out the ghost variables $X^{a}$ and to keep the ghost momenta $\Pi_{a}$, which yields:
\begin{widetext}
\begin{eqnarray}
\label{EvolutionOperatorPathIntBRST1}
U( W(t_{1}), W(t_{2}) ) = \int \limits_{W(t_{1})}^{W(t_{2})}
\mathcal{D} W(t) \: \mathcal{D} \Pi(t) \: \mathcal{D} \chi(t) \: \mathcal{D} (nonminimal) \: \nonumber \\
\delta \left(\dot{\Pi} \right) \delta \left(\dot{\lambda}^{0 a}_{0} \right)
\delta \left(\dot{\chi}^{k} - \rho^{0 k}_{1} \right) \delta \left(\dot{\rho}^{0 a}_{0} \right)
\delta \left(\dot{\rho}^{0 k}_{1} \right)
\nonumber \\
\exp{ \left( -i \int \limits_{t_{1}}^{t_{2}} d t \left( \frac{1}{2 g^{2}_{YM} L} \:
\tr \left(W^{-1} \dot{W} + \lambda \right)^{2} + \frac{g^{2}_{YM} L}{2} \: \chi^{k} h_{k}^{a b} \Pi_{a} \Pi_{b} -  \frac{g^{2}_{YM} L}{2} \: \Pi_{a} \rho^{0 a}_{0}
\right)
\right)}
\end{eqnarray}
\end{widetext}
where $\lambda = \lambda^{0 a}_{0} Q_{a}^{\alpha} L_{\alpha}$ is the matrix-valued lagrange multiplier and the factors $\frac{g^{2}_{YM} L}{2}$ were restored again. The constraints $\dot{\Pi}=0$, $\dot{\lambda}^{0 a}_{0}=0$, $\dot{\chi}^{k} = \rho^{0 k}_{1}$, $\dot{\rho}^{0 a}_{0}=0$, $\dot{\rho}^{0 k}_{1}=0$ imply that the only variable which depends on time is $W(t)$, while for all other variables the path integral reduces to the ordinary one.

\section{Conclusions}

 In this paper BRST quantization of the one-dimensional matrix model (\ref{2DYMSphereMMLagrangian}) with the constraints (\ref{2DYMTorusMMConstraint}) was performed. The problem was reformulated in terms of classical and quantum mechanics on the group manifold, and BRST invariance requirement was imposed in the hamiltonian formalism. Classical and quantum BRST-invariant hamiltonians (\ref{BRSTHamiltonianClassic}) and (\ref{BRSTHamiltonianQuantum}) were constructed. The kernel of the evolution operator between physical states was expressed in terms of BRST path integral (\ref{EvolutionOperatorPathIntBRST}) and (\ref{EvolutionOperatorPathIntBRST1}).

In order to carry out similar analysis directly in the lagrangian formalism, one can use the antifield formalism developed by Batalin, Fradkin and Vilkovisky \cite{Batalin:83:1} (also known as the lagrangian BRST quantization) which is equivalent to BRST quantization \cite{HenneauxQuantGauge}. As in the antifield formalism one needs exactly as many auxiliary variables as in the BRST formalism \cite{HenneauxQuantGauge, Batalin:83:1}, the complexity of the problem is the same for both methods. Antifield formalism can be also suitable for zero-dimensional matrix models which are of special interest in quantum gravity and string theories \cite{Marino:04:1, MakeenkoGaugeMethods}.

It is important to note that as the gauge group was not specified explicitly, one can apply the above analysis to such non-simple gauge groups as $\left(SU(N) \right)^{M}$, i.e. to the models with several matrix-valued variables, where the reduced phase space can not be constructed by diagonalization as in \cite{Minahan:93:1, Douglas:93:1}. It seems that infinite-dimensional groups such as $SDiff(\mathcal{M})$ could also be considered \cite{Hoppe:88, Pope:90:1}, which could be interesting in the context of correspondence between large $N$ two-dimensional Yang-Mills theories and string theories \cite{Gross:93:1, Gross:93:2, Vafa:04},  however, this possibility requires more detailed investigation of topological features of these groups.

\begin{acknowledgments}
I would like to thank Dr. Hossein Sadeghpour for his kind hospitality at ITAMP (Harvard University). I am also grateful to the members of the ITEP lattice group (ITEP, Moscow), especially to Dr. M. I. Polikarpov and E. Luschevskaya, for their hospitality, support and stimulating discussions. This work was partially supported by the U.S. National Science Foundation through a grant for the Institute for Theoretical Atomic, Molecular and Optical Physics at Harvard University and Smithsonian Astrophysical Observatory.
\end{acknowledgments}

\appendix

\section{Solution of the equations for $h^{a b}_{k}$}
\label{sec:ClosednessSolution}

In order to solve the equations (\ref{ClosednessCondition}) suppose that $h^{a b}_{k} u_{a \: l} = 0$, i.e. that $h^{a b}_{k}$ is an antisymmetric tensor in the subspace of the Lie algebra spanned on $Q^{a}_{\alpha}$. After such assumption the $2$-form $h_{k \: \alpha \beta}$ can be used instead of $h^{a b}_{k}$ without loosing any components of $h^{a b}_{k}$:
\begin{equation}
\label{hFormDef}
h_{k \: \alpha \beta} = h_{k}^{a b} Q_{a \: \alpha} Q_{b \: \beta}
\end{equation}
where $Q_{b \: \alpha} = g_{\alpha \beta} Q^{\beta}_{b}$. The first equation in (\ref{ClosednessCondition}) can be rewritten as:
\begin{eqnarray}
\label{ClosednessCondition1}
\partial_{\beta} u^{a}_{k} + h^{a b}_{k} Q_{b \: \beta} = 0
\end{eqnarray}
After the redefinition (\ref{hFormDef}) the equation (\ref{ClosednessCondition1}) directly yields the $2$-form $h_{k \: \alpha \beta}$:
\begin{eqnarray}
\label{ClosednessConditionDiffGeomPrelim}
Q_{a \: \alpha} \partial_{\beta} u^{a}_{k} + h^{a b}_{k} Q_{a \: \alpha} Q_{b \: \beta} = 0
\nonumber \\
 - u^{a}_{k} \nabla_{\beta} Q_{a \: \alpha} + h^{a b}_{k} Q_{a \: \alpha} Q_{b \: \beta} = 0
\nonumber \\
 1/2 \: u^{a}_{k} \partial_{[ \alpha |} Q_{a \: | \beta ] } + h_{k \: \alpha \beta} = 0
\end{eqnarray}
where $\nabla_{\alpha}$ is the covariant derivative constructed from the metric $g_{\alpha \beta}$. As $Q_{a \: \beta}$ is the Killing vector, $\nabla_{( \alpha |} Q_{a \: | \beta )}$ = 0, and $\nabla_{\beta} Q_{a \: \alpha} = 1/2 \: \nabla_{(\beta |} Q_{a \: | \alpha )} + 1/2 \: \nabla_{[ \beta |} Q_{a \: | \alpha ]} = 1/2 \: \nabla_{[ \beta | } Q_{a \: | \alpha ]} = 1/2 \: \partial_{[ \beta |} Q_{a \: | \alpha ]}$.
The last equation in (\ref{ClosednessConditionDiffGeomPrelim}) can be rewritten in the compact form as:
\begin{eqnarray}
\label{ClosednessConditionDiffGeom}
h_{k} =  - 1/2 \: u^{a}_{k} d Q_{a}
\end{eqnarray}
where $d$ is the external derivative on the group manifold.

Using the commutation relations (\ref{CommutationRelationsClassic}) the second equation in (\ref{ClosednessCondition}) can be reformulated as $\delta_{Q_{c}} h_{k} = 0$, where $\delta_{Q_{c}}$ is the Lie derivative along the vector field $Q^{\alpha}_{c}$, i.e. the 2-form $h_{k}$ should be invariant w.r.t. the diffeomorphisms generated by $Q_{a}^{\alpha}$. The Lie derivative of $\delta_{Q_{c}} h_{k}$ can be calculated using the definition (\ref{ClosednessConditionDiffGeom}):
\begin{eqnarray}
\label{hInvarianceProof}
\delta_{Q_{c}} h_{k} =  - 1/2 \:  \delta_{Q_{c}} \left( u^{a}_{k} d Q_{a} \right) =
\nonumber \\ =
 - 1/2 \:  \delta_{Q_{c}} u^{a}_{k}  d Q_{a}  - 1/2 \:  u^{a}_{k} d \delta_{Q_{c}} Q_{a} =
\nonumber \\ =
 - 1/2 \:  C_{f c}^{a} u^{f}_{k} d Q_{a}  + 1/2 \: u^{a}_{k} C_{a c}^{f} d Q_{f} = 0
\end{eqnarray}
where the commutation relations (\ref{CommutationRelationsClassic}), (\ref{NilpotencyCondition}) and the commutativity of external and Lie derivatives were used.

To finally prove the existence of the solution of (\ref{ClosednessCondition}) it is sufficient to demonstrate that $h_{k \: \alpha \beta}$ indeed can be represented in the form $h^{a b}_{k} Q_{a \: \alpha} Q_{b \: \beta}$, which is equivalent to $h_{k \: \alpha \beta} u^{\beta}_{l} = 0$. The external derivative $ \left( d Q_{a} \right)_{\alpha \beta} = \partial_{[ \alpha |} Q_{a \: | \beta ]}$ can be found using the fact that $L^{a}_{\alpha}$ and $-R^{a}_{\alpha}$ are non-Abelian full derivatives and therefore have vanishing Yang-Mills curvature forms:
\begin{eqnarray}
\label{dQExplicit}
\partial_{[ \alpha} Q^{a}_{\beta ]} = \partial_{[ \alpha} L^{a}_{\beta ]} - \partial_{[ \alpha} R^{a}_{\beta ]} =
\\ \nonumber
- C^{a}_{b c} L^{b}_{\alpha} L^{c}_{\beta} - C^{a}_{b c} R^{b}_{\alpha} R^{c}_{\beta}
\end{eqnarray}
Now $h_{k \: \alpha \beta} u^{\beta}_{k}$ can be calculated:
\begin{eqnarray}
\label{huExplicit}
h_{k \: \alpha \beta} u^{\beta}_{l} =  - 1/2 \:  u_{a \: k} \partial_{[ \alpha} Q^{a}_{\beta ]} u^{\beta}_{l} =
\nonumber \\ =
  + 1/2 \:  u_{a \: k} C^{a}_{b c} L^{b}_{\alpha} L^{c}_{\beta} u^{\beta}_{l} +  1/2 \:  u_{a \: k} C^{a}_{b c} R^{b}_{\alpha} R^{c}_{\beta} u^{\beta}_{l} =
\nonumber \\ =
 - 1/2 \:  u_{a \: k} C^{a}_{b c} L^{b}_{\alpha} u^{c}_{l}  + 1/2 \:  u_{a \: k} C^{a}_{b c} R^{b}_{\alpha} u^{c}_{l} = 0
\end{eqnarray}
where the identities $u^{\alpha}_{k} L^{a}_{\alpha} = u^{\alpha}_{k} R^{a}_{\alpha}$ and $u^{a}_{k} u^{b}_{l} C_{a b}^{c} = 0$ were used. Thus the 2-form $h_{k \: \alpha \beta}$ was found. As $h^{a b}_{k}$ was assumed to lie in the subspace spanned on $Q^{a}_{\alpha}$, it  can be extracted from $h_{k \: \alpha \beta}$ by contracting it with some vector fields $S^{\alpha \: a}$ which are pseudoinverse to $Q^{a}_{\alpha}$, i.e. $S^{\alpha}_{a} Q^{b}_{\alpha} = P^{b}_{a}$, where $P^{b}_{a}$ is the projector on the subspace in the Lie group spanned on $Q^{a}_{\alpha}$. The vector fields $S^{\alpha \: a}$ are explicitly constructed in Appendix \ref{sec:SUNPolar} for $SU(N)$ group. Thus  the coefficients $h^{a b}_{k}$ which enter the BRST-invariant hamiltonian (\ref{BRSTHamiltonianClassic}) have been constructed.

\section{Angular and radial coordinates on $SU(N)$ groups}
\label{sec:SUNPolar}

In this Appendix special coordinates on the group manifold are constructed and the existence of the vector fields $u^{\alpha}_{k}$ which satisfy the commutation relations (\ref{NilpotencyCondition}) and (\ref{uCommutation}) is demonstrated.

Any $SU(N)$ element can be represented as:
\begin{eqnarray}
\label{SUDiagonalization}
U = V D V^{-1}
\end{eqnarray}
where $D$ is a diagonal $SU(N)$ matrix and $V$ is some unitary matrix. The matrix $D$ may be represented as $D = \exp{\left( i \phi^{k} T_{k} \right)}$, where $T_{k}$ are the elements of the Cartan subalgebra of $su(N)$ Lie algebra, i.e. diagonal traceless hermitian matrices. $\phi^{k}$ can be chosen as the first $N - 1$ coordinates. The matrices $V$ should depend on the other $N^{2} - N$ coordinates only, which will be denoted as $\theta^{\mu}$ and labelled with the indices $\mu, \nu, \ldots$ from the second half of the greek alphabet. Thus the coordinates on the group are split into the radial part $\phi^{k}$ and the angular part $\theta^{\mu}$: $x^{\alpha} = \left( \phi^{k}, \: \theta^{\mu} \right)$.

For the right- and left- invariant matrix-valued 1-forms $L_{\alpha}$ and $R_{\alpha}$ (\ref{LRFormsMatrix}) one obtains from (\ref{SUDiagonalization}):
\begin{eqnarray}
\label{SULR1Forms}
\partial_{\alpha} U = i S_{\alpha} U - i U S_{\alpha} + V \partial_{\alpha} D V^{-1} \nonumber \\
R_{\alpha} = - i U^{-1} \partial_{\alpha} U = U^{-1} S_{\alpha} U - S_{\alpha} + V A_{\alpha} V^{-1}
\nonumber \\
L_{\alpha} = - i \partial_{\alpha} U U^{-1} =  S_{\alpha} - U S_{\alpha} U^{-1} + V A_{\alpha} V^{-1}
\end{eqnarray}
where $S_{\alpha} = -i \partial_{\alpha} V V^{-1}$, $S^{a}_{\alpha} = \tr \left( T^{a}  S_{\alpha} \right)$ and $A_{\alpha} = -i \partial_{\alpha} D D^{-1}$. By definition $S_{\alpha} = \left(0, S_{\mu} \right)$ and $A_{\alpha} = \left( A_{k} , 0 \right)$. Note that the representation (\ref{SUDiagonalization}) is not unique, since one can always change $V \rightarrow V G$, where $G$ is a diagonal unitary matrix. Under such "gauge transformations" $S_{\alpha}$ transforms as:
\begin{eqnarray}
\label{SGaugeTransform}
S_{\alpha} \rightarrow S_{\alpha} - i V \partial_{\alpha} G G^{-1} V^{-1}
\end{eqnarray}

The metric tensor $g_{\alpha \beta}$ is block-diagonal w.r.t. the coordinates $\phi^{k}$ and $\theta^{\mu}$:
\begin{eqnarray}
\label{MetricTensor}
g_{\alpha \beta} = \tr \left( L_{\alpha} L_{\beta} \right) = \nonumber \\ =
\tr \left( 2 S_{\alpha} S_{\beta} - U^{-1} S_{\alpha} U S_{\beta} -
U^{-1} S_{\beta} U S_{\alpha} + A_{\alpha} A_{\beta}
 \right) \nonumber \\
 g_{\mu \nu} =  \tr \left( 2 S_{\mu} S_{\nu} - U^{-1} S_{\mu} U S_{\nu} -
U^{-1} S_{\mu} U S_{\nu} \right)
\nonumber \\
g_{k l} =  \tr \left( A_{k} A_{l} \right), \quad g_{\mu k} = 0 \quad
\end{eqnarray}
The 1-form $Q_{\alpha} = Q^{a}_{\alpha} T_{a} = L_{\alpha} - R_{\alpha}$ has only $\theta^{\mu}$ components:
\begin{eqnarray}
Q_{\mu} = 2 S_{\mu} - U^{-1} S_{\mu} U - U S_{\mu} U^{-1}
\end{eqnarray}
The angular block of the metric can be written as $g_{\mu \nu} = \tr \left( Q_{\mu} S_{\nu} \right)$, therefore:
\begin{eqnarray}
\label{MetricAngular}
\tr \left( Q_{\mu} S^{\nu} \right) = Q^{a}_{\mu} S_{a}^{\nu} = \delta^{\nu}_{\mu}, \quad
S^{a}_{\mu} Q^{\mu}_{c} = P^{a}_{c}
\end{eqnarray}
where $P^{a}_{c}$ is the projector on the subspace in the Lie algebra spanned on $Q^{a}_{\mu}$. The transformations (\ref{SGaugeTransform}) does not change the angular part (\ref{MetricAngular}) of the metric (\ref{MetricTensor}) by definition of $Q^{a}_{\alpha}$.

Let the vector fields $u_{k}^{\alpha}$ be proportional to the differentials of the radial coordinates $\phi^{k}$, so that $u^{\alpha}_{k} \partial_{\alpha} = \partial_{k}$. Such vector fields commute among each other by definition. Lie bracket of the vector fields $u^{\alpha}_{k}$ and $Q^{\alpha}_{a}$ is equal to $\partial_{k} Q^{\mu}_{a}$, therefore the the equation (\ref{NilpotencyCondition}) can be now rewritten as $\partial_{k} \left( g^{\mu \nu} Q_{\nu}^{a} \right) = 0$, or, more explicitly:
\begin{eqnarray}
\label{NilpotencyCoordinateDependent}
\partial_{k} Q_{\mu} = \partial_{k} g_{\mu \nu} g^{\nu \sigma} Q_{\sigma}
\end{eqnarray}
The equation (\ref{NilpotencyCoordinateDependent}) is actually an identity which follows from the expression (\ref{MetricAngular}) for the angular block of the metric. Thus the vector fields $u_{k}^{\alpha}$ which satisfy the commutation relations (\ref{NilpotencyCondition}) and (\ref{uCommutation}) have been constructed.


\end{document}